\documentclass[12pt]{article}
\usepackage{bbm, amsfonts}
\normalbaselines 
\begin{document} 
\begin{center}
\vspace*{1.0cm}

{\LARGE{\bf Anomalies and Star Products}} 

\vskip 1.5cm

{\large {\bf H. R\"omer and C. Pauf\/ler}} 

\vskip 0.5 cm 

Fakult\"at f\"ur Physik\\ 
Albert-Ludwigs-Universit\"at Freiburg im Breisgau\\ 
Hermann-Herder-Stra\ss e 3\\ 
D 79102 Freiburg i. Br.\\
Germany

\end{center}

\vspace{1 cm}

\begin{abstract}
A formulation of anomalies in terms of star products is suggested which promises 
insight from an alternative and unifying point of view.
\end{abstract}

\vspace{1 cm} 

\section{Overview}

Anomalies correspond to the fact that sometimes symmetries of a classical
theory cannot be implemented in the corresponding quantum theory. The study of
field theoretical anomalies has a long history. They were originally observed as 
unexpected terms in the commutators of quantum currents among each others
(Schwinger terms) or with products of field operators (anomalous Ward
identities). Much insight was subsequently obtained in the nature of anomalies
by topological and cohomological methods.

One line of development lead to a description of gauge anomalies as local
cocycles on the gauge group expressed in terms of field operators. Relating
anomalies to the lack of invariance of the determinant of operators of Dirac
type yielded an understanding in terms of characteristic classes of determinant
bundles over the space of connections.

These characteristic classes are all obtainable from first quantization of a
Dirac type particle in an external classical gauge field, either in a
Lagrangian or in a Hamiltonian formulation. Sometimes it is also appropriate to
consider bundles of Fock spaces over the space of connections
(\cite{CaMiMu97b,Seg85}). 
In any case, the
information on anomalies is related to the bundles of null spaces of families of 
Dirac operators. The anomalies are traces of products of inverse powers of the
Dirac operator $D$ and certain vertex operators, just those expressions also
obtainable from one-loop Feynman integrals, where $D^{-1}$ corresponds to the
propagator of the Dirac field.

In yet another language, the anomalies are obtained from the symbol calculus of
differential operators on the Dirac bundle. Further insight is gained from
non-commutative geometry. Starting from the algebra of differential operators on
the Dirac bundle, non commutative characteristic classes can be found, which are 
just the non commutative counterpart of the anomalies and which, evaluated on
suitable non-commutative cycles, just give the characteristic numbers.

These results can also be looked at from the point of view of star products or
deformation quantization (\cite{BFFLS77,BFFLS78,BCG97}). 
The algebra $\mathfrak{A}$ of symbols of differential
operators on the Dirac bundle can be considered as the algebra of observables of 
a system consisting of a Dirac particle in an external gauge field. The symbol
calculus in this algebra is thus a special example of a star product on
$\mathfrak A$. More precisely, one has to go over to the algebra
$\mathfrak A[[\lambda]]$ of formal power series in an
indeterminate $\lambda$ (eventually to be substituted by $\hbar$) with
coefficients in $\mathfrak A$.

This restriction to formal power series is by no means only a shortcoming, but
it also has definite advantages.

It opens the way to a concentration on a better understanding of algebraic
aspects of the quantization procedure. In particular, there is the powerful notion
of equivalence of star products that are related, for example, by changing ordering
prescriptions in the quantization procedure. Anomalies, anyhow, are always of
low order in $\hbar$ and should be treatable without any loss of information in
terms of formal power series in $\lambda$. In the language of star product
quantization, anomalies are objects in the non-commutative geometry of the
algebra $\mathfrak A[[\lambda]]$ endowed with the star product. The trace functional
on $\mathfrak A[[\lambda]]$, applied to non-commutative characteristic classes on
$\mathfrak A[[\lambda]]$  give characteristic numbers. Uniqueness theorems on the
trace functional and non commutative cohomologies can successfully be applied. The
theory of star products on finite dimensional phase spaces like the phase space 
of the Dirac system is well defined. Anomalies are expected also to arise as
imperfections in invariance properties of the star product.

\section{A mechanical example}
Let $\mathcal M$ be a classical phase space of finite dimension. Let a group
$\mathcal G$ with Lie algebra $\mathfrak g$ act on $\mathcal M$ in a Hamiltonian 
way with equivariant momentum map (\cite{AM85})
\[
\mathcal J:\mathcal M\rightarrow\mathfrak g^\ast,
\]
such that for every $\xi\in \mathfrak g$, $f\in C^\infty(\mathcal M)$
\[
L_\xi f=-\{f,\mathcal J_\xi\} \quad \textrm{and} \quad 
\{\mathcal J_\xi,\mathcal J_\eta\}=\mathcal J_{[\xi,\eta]},
\]
where ${\mathcal J}_\xi(p)=\langle {\mathcal J}(p),\xi\rangle$.

A star product $\star$ on $C^\infty(\mathcal M)[[\lambda]]$ is called (see,
e.g., \cite{ACMP83})
\begin{enumerate}
\item \label{covariance}
covariant, if 
$[\mathcal J_\xi,\mathcal J_\eta]_\star=
\mathcal J_\xi\star\mathcal J_\eta
-\mathcal J_\eta\star\mathcal J_\xi=\textrm{i}\lambda\mathcal J_{[\xi,\eta]}$,
\item \label{invariance}
invariant, if
$\{\mathcal J_\xi,f\star g\}=\{\mathcal J_\xi,f\}\star g
+f\star\{\mathcal J_\xi,g\}$,
\item \label{strong-invariance}
strongly invariant, if
$\mathcal J_\xi\star f- f\star\mathcal J_\xi=\textrm{i}\lambda\{\mathcal J_\xi,f\}$.
\end{enumerate}
Clearly, the last property implies the two others. The second equation is the
infinitesimal version of the requirement of invariance under canonical
transformations $\Phi$:
\[
\Phi^\ast(f\star g)=\Phi^\ast f\star \Phi^\ast g.
\]
Schwinger terms should correspond to a violation of the covariance condition
(\ref{covariance}.), whereas anomalous terms in Ward identities are expected to
reflect 
themselves in a lack of strong invariance (\ref{strong-invariance}.).

The group of all classical canonical transformations, for example, cannot be
implemented in quantum theory without anomalies. Indeed, for this large group,
all observables are momentum maps, conditions (\ref{covariance}.) and
(\ref{invariance}.) become identical and  
the impossibility of a star product to transform Poisson brackets to star
commutators directly follows from the theorem of Groenewald and van Hove. These
anomalies are the reason for the notorious incompabilities of quantization
procedures and canonical transformations. If conditions (\ref{covariance}.),
(\ref{invariance}.) or (\ref{strong-invariance}.) are not fulfilled, one may try
to save the situation by adding non leading terms to $\mathcal J_\xi$,
\[
\mathbbm J_\xi=\mathcal J_\xi+\lambda\cdot(\ldots),
\]
such that $\mathbbm J_\xi$, the so-called quantum momentum map (\cite{Xu98}),
shows the above properties. 

For example, quantum covariance means
\[
\mathbbm J_\xi\star\mathbbm J_\eta-\mathbbm J_\eta\star\mathbbm J_\xi
=\textrm{i}\lambda\mathbbm J_{[\xi,\eta]}.
\]
One easily convinces oneself, that covariance holds for a star product $\star$
if and only if it holds for all equivalent star products. Hence, the
transformation 
\[
T:\mathcal J_\xi\rightarrow\mathbbm J_\xi
\]
rendering a non-covariant $\mathcal J_\xi$ covariant cannot be an equivalence
transformation but must depend on $\xi$. The redefinition of $\mathcal J_\xi$ is 
ad hoc rather than global.

As an illustration we can consider a mechanical system in one space
dimension. Galilei invariant systems are not appropriate, because there is no
equivariant momentum map. The generators of translations and boosts $p$ and $q$
commute
as Lie algebra elements, but as generators of canonical transformations we have
$\{p,q\}=1$. The Galilei group can be realized by canonical transformations only 
after central extension.

The Poincar\'e group does not have this defect. The Ruijsenaars-Schneider model
(\cite{RuijSch86})
is a mechanical system in one space dimension with Poincar\'e invariance and non 
trivial interaction.

Let us denote the generators of time and space translations by $\mathfrak h$ and 
$\mathfrak p$ and the boost generator by $\mathfrak b$. In the Lie algebra of
the Poincar\'e group we have 
\[
[\mathfrak h,\mathfrak p]=0,\quad [\mathfrak b,\mathfrak h]=\mathfrak p,\quad
[\mathfrak b,\mathfrak p]=\frac{1}{c^2}\mathfrak h.
\]
In the Ruijsenaars-Schneider model, 
the same relations hold,
\[
\{\mathfrak h,\mathfrak p\}=0,\quad \{\mathfrak b,\mathfrak h\}=\mathfrak p,\quad
\{\mathfrak b,\mathfrak p\}=\frac{1}{c^2}\mathfrak h,
\]
where $\mathfrak h$, $\mathfrak p$, and $\mathfrak b$ now are certain
functions over phase space. 
This is not in contradiction with the no-interaction theorem, because the
translations are realised in the conventional way only to order $\frac1{c^2}$.
Moreover, in a canonical quantization one has
\[
[\mathbbm H,\mathbbm P]=0,\quad 
[\mathbbm B,\mathbbm H]=-\frac{\textrm{i}}\hbar\mathbbm P, \quad
[\mathbbm B,\mathbbm P]=-\frac{\textrm{i}}{\hbar c^2}\mathbbm H,
\]
or 
(the factor $-\textrm i$ can be absorbed in the formal parameter $\lambda$)
\[
[\hat{\mathfrak h},\hat{\mathfrak p}]_\star=0,
\quad [\hat{\mathfrak b},\hat{\mathfrak h}]_\star=\hat{\mathfrak p},
\quad
[\hat{\mathfrak b},\hat{\mathfrak p}]_\star=\frac{1}{c^2}\hat{\mathfrak h}
\]
for the standard ordered star product $\star$.
Here, $\hat{\mathfrak h}$, $\hat{\mathfrak p}$, and $\hat{\mathfrak b}$ 
differ from the corresponding classical
functions only in higher orders of $\lambda$.
In addition the classical Ruijsenaars-Schneider system is completely
integrable. It has a series of conserved quantities $\mathcal J_k$ in
involution,
\[
\{\mathcal J_k,\mathcal J_l\}=0.
\]
In the quantized theory one has, as shown by Ruijsenaars (\cite{Ruij87}), 
that a modification
$\mathcal J_k\rightarrow\mathbbm J_k$ is possible, such that quantum integrability 
\[
[\mathbbm J_k,\mathbbm J_l]_\star=0
\]
holds, where $\star$ again is the standard ordered star product. 

\section{Field theory}

One would, of course, like to apply the star product formalism directly to field
theory, not only to first quantization. Unfortunately, for this case, the theory 
of star products is not yet fully developed.

One reason lies in the fact that even for the classical algebra of fields
products like $\varphi_1(x_1)\varphi_2(x_2)$ are in general undefined for
$x_1=x_2$, this deficiency beeing even more serious for quantum fields. The normal
way out is to consider (quantum) fields as (operator valued) distributions. For
objects 
\[
  \varphi_f
  =\int \,d^nx_1\ldots
  d^nx_r\,\varphi(x_1)\cdots\varphi(x_r)\,f(x_1,\ldots,x_r)
\]
with suitable test  
functions $f$, products 
are now defined (\cite{DF00} and references therein). 
For the same reason, one expects star products
$\varphi_f\star\varphi_g$ to be well defined.

In quantum field theory, one normally concentrates on expectation values like
\[
w(x_1,\ldots,x_n)=\langle0|\varphi(x_1)\ldots\varphi(x_n)|0\rangle.
\]
This should also be a promising strategy for field theory in star product
formulation. One can interprete the distribution
\begin{eqnarray*}
w(x_1,\ldots,x_n)&=\langle0|\varphi(x_1)\ldots\varphi(x_n)|0\rangle\\
&=\mu\left(\varphi(x_1)\star\ldots\star\varphi(x_n)\right)
\end{eqnarray*}
as the value of a state on the algebra $\mathcal A[[\lambda]]$ of classical
fields endowed with a star product. In fact, $w$ even resembles a trace
functional on $\mathcal A[[\lambda]]$. $w$ has to be interpreted as a formal
series in $\lambda$, whose coefficients are just the quasi-classical expansion of 
the quantum correlation functions. This will result in considerable technical
simplifications as compared to quantum field theory. 
The star product algebra can now be reconstructed by a generalized GNS
construction in the sense of the method introduced by M. Bordemann and
S. Waldmann (\cite{BW98}). Recently, part of this program has been implemented by
K. Fredenhagen and M. D\"utsch in \cite{DF00}.

For the discussion of anomalies, the momentum map $\mathcal J_\xi$ has to be
replaced by the conserved current $\mathcal J_\xi^\mu$. Adding higher
terms to find a quantum momentum map corresponds to the need of
regularization. Indeed, by using the canonical (anti)commutation relations
the classical expression for normal ordered currents is
found formally only to order $\hbar$.

Conditions similar to (\ref{covariance}.), (\ref{invariance}.), and
(\ref{strong-invariance}.) can be formulated, and their violation will be largely 
analogous to anomalous Ward identities in quantum field theory but in the sense 
of formal power series. Violations of the above conditions should be
characterized cohomologically.

There should be a formulation of anomalies in terms of the non-commutative
geometry in the field algebra $\mathcal A[[\lambda]]$ and not just in the
algebra $\mathcal A_D[[\lambda]]$ of a Dirac particle.

The Dirac Operator plays a fundamental r\^ole in the non-commutative geometry of 
$\mathcal A_D[[\lambda]]$. $D$ can also be interpreted as a derivation in
$\mathcal A$, which can also be written in the form 
\[
\mathcal A[[\lambda]]\ni a\mapsto[Q,a]_\star\in\mathcal A[[\lambda]],
\]
with
\[
Q=\int\,d^nx\,\bar\varphi D\varphi.
\]
A more ambitious task would be to formulate anomalies as algebraic properties on 
the algebra $\mathcal B[[\lambda]]$ of observables of a field theory rather than
on the larger field algebra $\mathcal A[[\lambda]]$. The Dirac operator will
still be a derivation in $\mathcal B[[\lambda]]$. Gauge anomalies primarily
concern non observable quantities, their effect should be hidden more deeply in
$\mathcal B[[\lambda]]$.

\section*{Acknowledgements} 

This note is dedicated to H. D. Doebner.\\
We thank M. Bordemann for many elucidative discussions. H. R. thanks
J. Mickelsson for hospitality at the Royal Technical Highschool Stockholm and
for useful discussions. 

The revised version was induced by learning about a paper by D\"utsch
and Fredenhagen (\cite{DF00}), in which important elements of 
the construction sketched here were
implemented in considerable detail. We also thank M. D\"utsch and 
K. Fredenhagen for
illuminating discussions.

\end{document}